# Experimental and computer simulation studies of the micelles formed by comb-like PEG-containing polymeric surfactants as potential enzyme scaffolds

O. Paiuk[1], A. Zaichenko[1], N. Mitina[1], J.Ilnytskyi[1,2*], T.Patsahan[2], S.Minko[3], Kh. Harhay[1], V. Garamus[4], K. Volianiuk[1]

[1]Lviv Polytechnic National University, Lviv, Ukraine
[2]Institute for Condensed Matter Physics of the National Academy of Sciences of Ukraine, Lviv, Ukraine
[3]Nanostructured Materials Lab, University of Georgia, Athens, GA, USA
[4]5 Helmholtz-Zentrum Geesthacht (HZG): Centre for Materials and Coastal Research, Max-Planck-Str, Geesthacht 21502, Germany

*email: iln@icmp.lviv.ua

**Abstract**

The industrial implementation of biofuel production from lignocellulosic biomass faces a number of economic obstacles. One of these is the cost of enzymes, typically used for cellulose hydrolysis. Nature provides some hints towards the efficiency of this process, exampled in natural enzyme complexes –cellulosomes, produced by some microorganisms. Therefore, many research groups target synthetic routes to mimic such cellulosomes with synthetic structures when many questions remain to be addressed: the optimal chemical structure and size of such synthetic scaffolds, their adsorption on the cellulosic biomass particles, combinations, and best practices arrangement of enzymes in the complex. In this work, polyethylene glycol (PEG) copolymers that form micelles and accommodate enzymes in the micellar structures are systematically studied using both experimental and computer simulation techniques. Preliminary results indicate that the micelles are efficient polymer - enzymes structures for cellulose hydrolysis. While the direct quantitative comparison between the real and model systems is not always straightforward, both approaches agree on the role of the molecular architecture of the copolymers on micelle formation and their structural characteristics.

## 1.Introduction

The lignocellulosic biomass is a huge natural resource for biofuel production, but industrial implementation is restricted by relatively high cost of this process. Its bottleneck is the hydrolysis stage, due to combination of factors[1-2]. During this stage, the plant cell wall polysaccharides are decomposed by enzymes into sugar monomers, with their fermentation being the following step in the biofuel production. The cost of enzymes counts for about half of biomass hydrolysis process, hence, their efficient use is the key for the efficiency of this technology overall.

The most widely used way to perform hydrolysis is to use the mixtures of free enzymes. In this way the process turns into rather stochastic, when the sequence of requires stages, such as cutting bonds of polysaccharides, cleaving their ends and subsequent split into glucose requires availability of enzymes of different type to be "at the right place at the right time". On contrary, a perfect "team work" of enzymes is demonstrated by the polymer-enzyme complexes, produced by certain bacterias, and since discovery of that, stimulates researchers to mimic such synergy by using the synthetic carriers with either random[3] or specific arrangement[4] of enzymes on a polymer scaffold. Such synthetic polymersomes involve may various types of molecular architectures including linear, branched of hyperbranched examples[5].

Besides plethora of molecular architectures, the polymersomes may differ in the physical-chemical mechanism of enzymes immobilization, including cases of both covalent[6] and non-covalent[7] (by means of van der Waals or electrostatic forces) attachments. The non-covalent attachment relates such host-guest systems to the micelles that are used in the systems of targeted drug delivery and in other applications[8]. In the case of micellar polymersomes, all the factors such as: micelle forming dynamics, micelle hosting ability, stability, size and shape of the polymer-enzyme complexes are of most interest and require detailed study.

The popular class of surfactants are the PEG-containing amphiphilic polymers[9-13], that are used in various fields of science and technology as surfactants and surface modifiers. Such polymers are widespread as carriers for drug delivery systems, binding and delivery of proteins, and antibodies[14,15]. The intra- and intermolecular interactions of PEG chains provide their self-assembling and formation of micelle-like structures in the solutions in a wide range of polarities and on surfaces[16,17]. Experimental study of rheological characteristics of water solution of PEG with molecular weight higher than 600 Da[18-21] confirmed formation of intermolecular aggregates due to formation of the hydrogen bonds between such molecules. Moreover, an availability of $-CH_2-CH_2-$ units in the PEG chains causes the possibility of self-assembling the PEG-containing molecules via hydrophobic interaction depending, evidently, on the separation between the PEG chains. It was shown that the assembling degree and morphology of the micelles depends on polymer PEG chain length, molecular weight, architectures and fine molecular structure[22].Nevertheless, the rheological and colloidal-chemical properties of comb-like copolymers containing side PEG chains of different lengths and arrangement of side chain groups along the polymer backbone are not yet studied in a required detail.

In this work we provide the results for the synthesis of various PEG-containing polymer molecules of a side-chain architecture, study their ability to micellize, examine a range of relevant properties of the micelles and try to relate these to the enhancement of the glucose production. Although, the gain in the latter, as compared to the case of free enzymes, in not breathtaking, this paves a way for the future search of similar molecular architectures with more efficient glucose production, which will be the topic of the following studies. For deeper understanding the mechanisms of micellization and micellar structure, we combine both the experimental and computer simulation studies, where the modeling is performed on a mesoscopic level using the dissipative particle dynamics approach. The outline of the paper is as follows. Section 2 contains detailed description of the synthetic protocols and the results of the experimental studies, computer simulation studies are given in section 3, and the conclusions are provided in section 3.

### 2.Synthetic protocols and the experimental results

Poly(ethylene glycol) methyl ether methacrylate, with the molecular weight of 475 Da (PEGMA475) and purity 97 %, was received from Sigma Aldrich (Milwaukee, WI, USA) and used as received. Peroxide derivative of isopropyl benzene isopropyl-3(4)-[1-(tert-butylperoxy)-1-methylethyl] benzene (monoperoxine, MP) was synthesized using the method describe in Ref. [20]. MP has following constants after vacuum distillation and drying under magnesium sulfate: $d_4^{20}$= 0.867 (lit. 0.867); $n_d^{20}$=1.448 (lit. 1.448). Butyl acrylate (BA), dimethyl maleate (DMM), N-vinylpyrrolidone (NVP) were received from Merk and purified by vacuum distillation.All other solvents and reagents were obtained from Aldrich (Milwaukee, WI, USA).

Comb-like PEG-containing macromolecules of various architectures were synthesized via radical homo- or copolymerization of PEGMA475 ([monomer]=1.0mol/l) initiated by AIBN ([AIBN]=0.06 mol/l) in the presence of MP ([MP]=0.25mol/l) as chain transfer agent. Reaction mixtures preparing for PEGMA polymerization are described below. For synthesis of poly(PEGMA)-MP polymer, AIBN was dissolved in dry 1.4-dioxane, then PEGMA and MP were added to AIBN solution. Synthesis of poly(PEGMA475-co-BA)-MP and poly(PEGMA475-co-DMM)-MP was performed as explained below.AIBN was dissolved in dry 1.4-dioxane, then PEGMA (Mn = 475 Da) and BA or DMM were added to the solution ([general monomer concentration]=1.0 mol/l). The molar ratio of PEGMA475:BA and PEGMA:DMM were ranged from 2.9:91.7% mol. to 70.8:29.2 % mol. and from 2.9:91.7 % mol. to 73.2:26.8 % mol., respectively. The reaction mixtures were loaded into calibrated dilatometers and purged with argon. Polymerization was carried out at 343K till monomer conversion of 60-70% has been reached, where the latter was controlled by using the dilatometric and gravimetric techniques. Resulting products were cooled down to the ambient temperature, dried and fractioned via dropped precipitation and the fractions were dried in vacuum till a constant weight has been reached. Poly(PEGMA475-co-NVP)-MP was synthesized via polymerization carried out at 343K till monomer conversion of 60-70%, where the latter was controlled by using the dilatometric and gravimetric techniques. Resulting products were cooled down to the ambient temperature, dried and fractioned via dropped precipitation and the fractions were dried in vacuum till a constant weight has been reached. To obtain copolymers with different contents of NVP units, a mixture of NVP: PEGMA monomers was used at the following molar ratios: 21.8:79.2 (for [PEGMA-NVP]-III-3); 75.6:24.4 (for [PEGMA-NVP]-III-2) and 48.4:51.6 (for [PEGMA-NVP]-III-1). Chemical structures of all polymer molecules are collected in Fig.1, whereas their molar weight fractions and molecular masses – in Table 1.

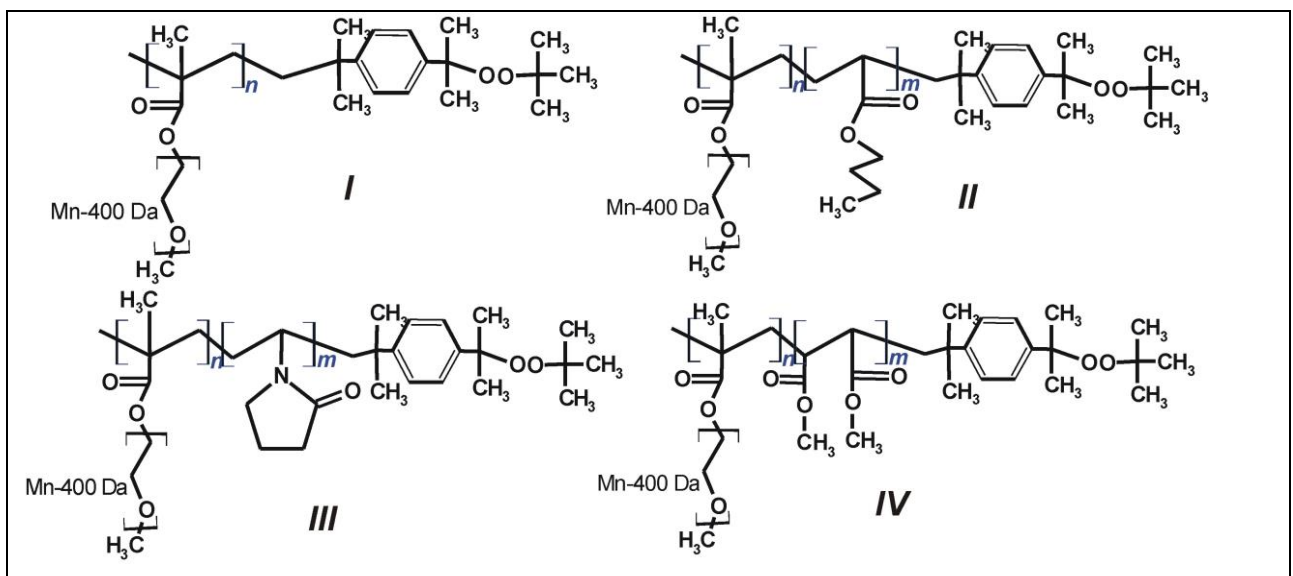

Fig.1. Chemical structures synthesized and discussed via computer simulations in current study. *I* stays for poly(PEGMA), *II* for poly(PEGMA-co-BA), *III* for poly(PEGMA-co-NVP) and *IV* for poly(PEGMA-co-DMM).

| copolymer | PEGMA fraction, % | BA fraction, % | DMM fraction, % | NVP fraction, % | MP fraction, % | polymer molar weight, kDa |
|---|---|---|---|---|---|---|
| [PEGMA] | 99.8 | - | - | - | 0.2 | 150.0 |
| [PEGMA-co-BA] | 85.7 | 14.0 | - | - | 0.3 | 105.0 |
| [PEGMA-co-DMM] | 85,2 | - | 14.0 | - | 0.8 | 25.0 |
| [PEGMA-co-NVP]-I | 84.0 | - | - | 15.4 | 0.6 | 45.2 |
| [PEGMA-co-NVP]-II | 93.05 | - | - | 6.5 | 0.45 | 59.5 |
| [PEGMA-co-NVP]-III-1 | 96.6 | - | - | 3.1 | 0.3 | 78.9 |

Tab.1. Molar mass fractions and molar weights of all polymers being synthesized and studied ecxperimentally.

As already pointed out in the Introduction, the main motivation for the synthesis and examination of the properties of this set of PEG-containing copolymers is their potential application as the cellulase carriers targeted on efficient cellulose hydrolysis. Here we show some preliminary results for that only, in a form of the comparison between the glucose production rate for cellulases immobilized within the micelles formed by these polymers vs the case of the mixture of free cellulases. These results are presented in Fig.2. More results and their deeper analysis, alongside with the results obtained for other type of polymers, will be given in a separate paper.

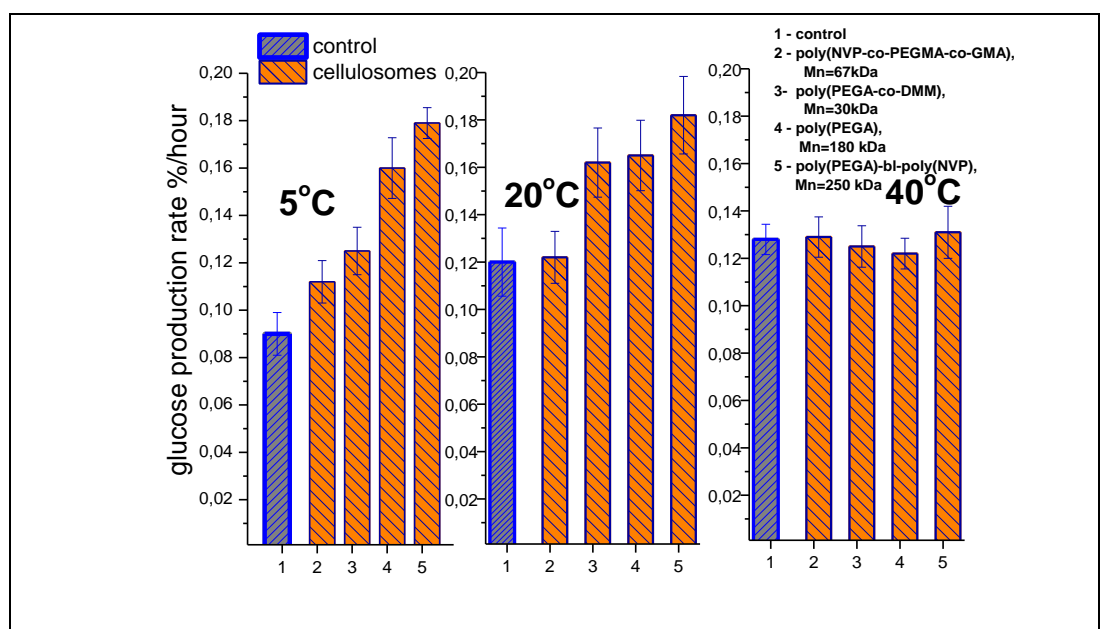

Fig.2. Glucose production rate as the result of hydrolysis performed by free cellulase (indicated as "Cellulase") and when cellulase is immobilized within a micelle formed of specific PEG-containing polymer, as indicated in the figure.

All these PEG-containing side chain polymers are known for their micellization abilities and are, to various extents, examined in relation to the targeted drug delivery [23]. In the current study we examine possible correlation between their micellization abilities and the efficiency of the cellulose hydrolysis. It might be expected, that the size, shape and internal structure of the micelles, serving as the hosts for cellulases, all play an important role in the process of hydrolysis. If so, then all these properties may be related to the experimentally observed differences in the glucose production rate shown in Fig.2.

Visualization of the micellar structures is performed via the TEM images are presented in Fig.3, from where one can observe the differences between their size and shape.

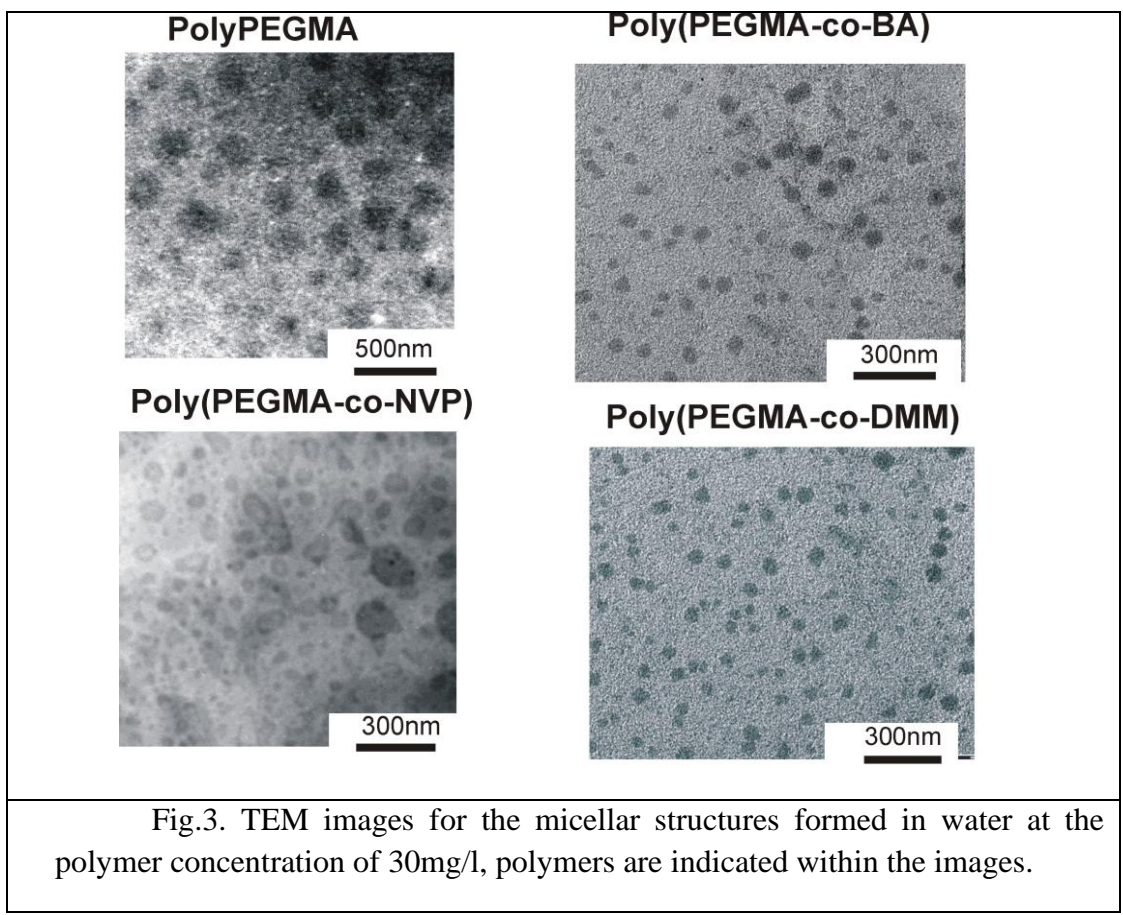

Fig.3. TEM images for the micellar structures formed in water at the polymer concentration of 30mg/l, polymers are indicated within the images.

| copolymer | $R_h$, nm | $CMC \cdot 10^4$, mol/l | $\sigma_{CMC}$, mN/m | $\Gamma_\infty \cdot 10^6$, mol/m$^2$ | $S_0$, Å$^2$ | NTU |
|---|---|---|---|---|---|---|
| [PEGMA] | 270 | 6.75 | 43.1 | 0.67 | 247 | 332 |
| [PEGMA-co-BA] | 160 | 8.26 | 39.3 | 0.92 | 180 | 101 |
| [PEGMA-co-DMM] | 152.5 | 23.3 | 46.7 | 1.11 | 150 | 3,7 |
| [PEGMA-co-NVP]-III-1 | 107.5 | 14.6 | 51.7 | 0.51 | 105 | 2.0 |
| [PEGMA-co- | 225 | 11.9 | 50.1 | 0.60 | 155 | 150 |
| NVP]-III-2 | | | | | | |
| [PEGMA-co-NVP]-III-3 | 240 | 9.75 | 49.3 | 0.60 | 180 | 210 |

Tab.2. Tab.2. Chemical colloidal characteristics of micelles are listed in Tab. 2. Here Rh is the average hydrodynamics radius measured via dynamic light scattering, CMC is the critical micelle concentration, is the surface tension, $\sigma_{CMC}$ the surface tension, $\Gamma_\infty$ is maximal adsorption at the water-air interface, $S_0$ is average minimum area per polymer molecule and NTU is turbidity of aqueous solution

The influence of the fine structures of the polymers on their solubility, surface activity and ability to form micelle-like self-organized assemblies in water solution is quantified via their respective colloidal-chemical characteristics, listed in Tab. 2, these are obtained used a number of experimental techniques specified below. Gel permeable chromatography (GPC) was used for polymeric molecular weight determination. Presented results were obtained by Waters 150C chromatograph with a builtin RI detector (Waters Corporation, Milford, USA), a Shodex 602 column (Kawasaki, Japan), the flow rate was varied (0.5 and 2.5 cm3 min–1). Dynamic light scattering(DLS) method was used for micellar sizes measurement by DynaProNanoStar(Wyatt Technology, Santa Barbara, USA) instruments and photon correlation spectra using the Non-InvasiveBack Scatter (NIBS) technology at 25 °C. For preparation of samples for DLS polymer was dissolved in bidistilled water (pH – 6.5 – 7.0) for 0.05 g·mL-1concentration and measured every sample by three-time. Solutions for DLS study were kept 24 h before the measurement. Hydrodynamic radii in this article presented by intensity of scattered light. TEM micrographs of the polymeric micelles were obtained with a transmission electron microscope JEM-200A (JEOL, Japan) at an accelerating voltage of 200 kV. The specimens for TEM were prepared by dropping 3 μL of sample solution onto a copper TEM grid (300 mesh) coated with thin, electron-transparent carbon film. After 1 min the solution was removed by touching the bottom of the grid with filtering paper (fast drying method). This fast removal of the solution was performed in order to minimize oversaturation artifacts during the drying process. Before the TEM observation, the samples were left to dry completely at room temperature.

Surface tension of polymeric surfactant solutions was studied by using the method of measurement of maximum bubble pressure[24] on Sensa Dyne bubble pressure tensiometer QC3000 (Gardco, USA). The device was calibrated using bidistilled water before the measurements. Polymeric solutions at different concentrations were prepared and keptfor at least 24 h before the measurement. It was assumed that equilibrium was reached if the surface tension variation was lesser 0.01mN·m$^{-1}$ during 10 min. PEG-containing polymers were dissolved in bidistilledwater, pH value was 6.5–7.0. The measurements were carried out in thethermostated Teflon cell at 25 °C.

The values of polymeric maximal adsorption ($\Gamma^\infty$, mol·m$^{-2}$)[25] were calculated using Eq. (1) and (2)[26] fromthe experimental plots ofisotherms of surface tension of the solutions.

$$\Gamma = -\frac{C}{RT\left(\frac{d\sigma}{d\lg C}\right)} \tag{1}$$

$$\left(\frac{C}{\Gamma}\right) = \left(\frac{C}{\Gamma^\infty}\right) + \left(\frac{\alpha}{\Gamma^\infty}\right) \tag{2}$$

in Eq. (1): - C (dσ/d lgC) is the slope of the curve of surface tension versus the logarithm of surfactant concentration; in Eq. (3): $1/\Gamma^\infty$ is the slope of the curve of dependence of C/Γ - C, and $\alpha/\Gamma^\infty$ is segment on the ordinate (α - constant Eq. (2)); R is an ideal gas constant, T is temperature. The average minimum area per molecule (S0,Å$^2$) is evaluated using the equation**Error! Bookmark not defined.**:

$$S0 = 1/(N_A \cdot \Gamma^\infty) \times 10^{20} \tag{3}$$

where $N_A$ is Avogadro's number.

Due to the tight arrangement of the PEG side chains along the backbone of poly(PEGMA)-MP, these have relatively rigid structure, and, therefore, are less surface active and form micelle-like assemblies at higher concentrations. Analysis of the results of colloidal-chemical study of other PEG containing comb-like surfactants confirms the determinative influence of the arrangement of side PEG chains on polymer dissolution and colloidal-chemical characteristics. One can see in Tab.2, that dilution of the PEGMA side chains by the hydrophobic DMM chains doesn't reduce polymer solubility as it was expected, moreover, the solubility of the poly(PEGMA-co-DMM)-MP in water increases significantly in comparison with poly(PEGMA)-MP. This is caused, in our opinion, by the enhancement of the accessibility of side PEG chains for water molecules and hydration. At the same time, this does not lead to the increase of the surface activity, micelle forming ability and impedes tight packing aggregates formed by poly(PEGMA-co-DMM)-MP molecules. As a result, self-assemblies formed by poly(PEGMA-co-DMM)-MP are characterized by smaller size (60-90 nm) in contrast to self-assemblies formed by poly(PEGMA)-MP molecules (from 100 nm to 500 nm) in water solution. The CMC value, surface tension and size of micelle-like assemblies formed by poly(PEGMA-co-BA)-MP in water solution are smaller than for poly(PEGMA)-MP due to, evidently, hydrophobic contribution of BA links in HLB value and increase of the backbone flexibility (Table 3, Fig. 5), however, for both comb-like polymers the sizes increase with the enhancement of polymer concentration in the solution and content of PEGMA units in molecules of poly(PEGMA-co-BA)-MP.

The images of the micelles from poly(PEGMA-co-BA)-MP molecules are obtained via the SAXS study and by TEM, these are shown in Fig.4. These indicate that linear clusters of these polymers form spherical assemblies at the increase of their aggregation number.

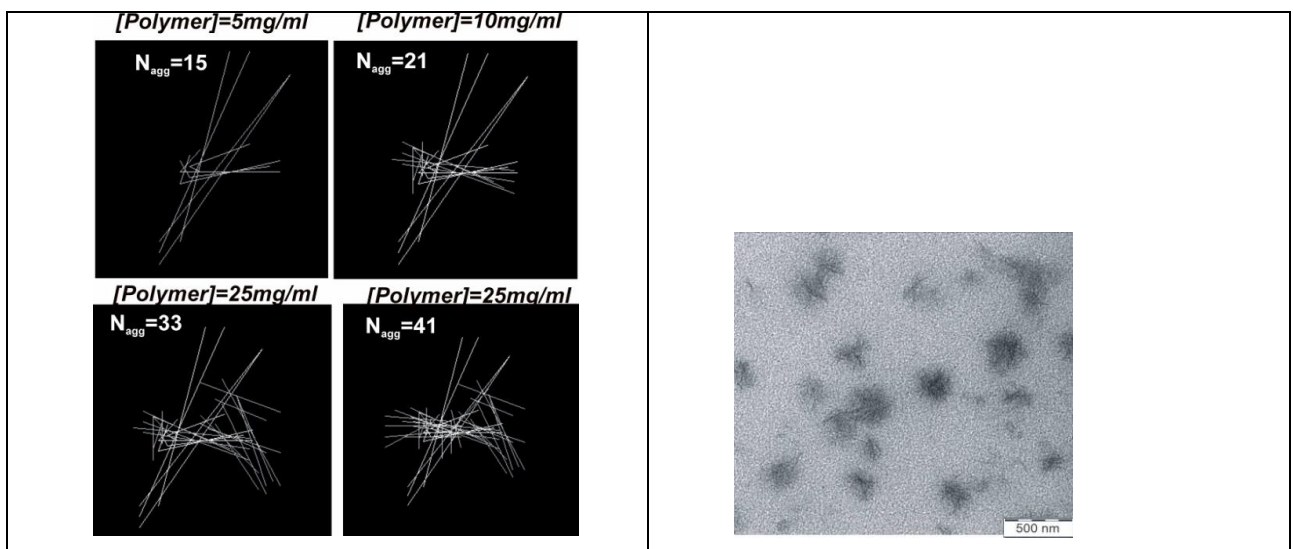

Fig. 4. 3Dmodelof the micelles formed by poly(PEGMA-co-BA)-MP at various polymer concentration in solution ([polymer] -= 5mg/ml (1); 10mg/ml (2); 25 mg/ml (3) and 50 mg/ml (4)) (on the l.h.s.), and the TEM imageofthe micellesformedbypoly(PEGMA475)-MP afterslowsampledrying onsubstrate (on the r.h.s.).

**3.Modeling, simulation technique and the results of simulations.**

Micellization in a water solutions of the polymer amphiphiles is modelled on a mesoscopic level, using the dissipative particle dynamics (DPD) simulation technique[27]. To this end, each such molecule was split into a set of relevant chemical groups, forming a LEGO-like structure of building blocks. In doing this, we followed closely the approach and the nomenclature therein, of recent works by the research group of Warren[28-29], which, in turn, is based on the concept of surface cite interaction points by Hunter[30]. The repeating groups of all amphiphile molecules being considered, are shown in Fig.5.

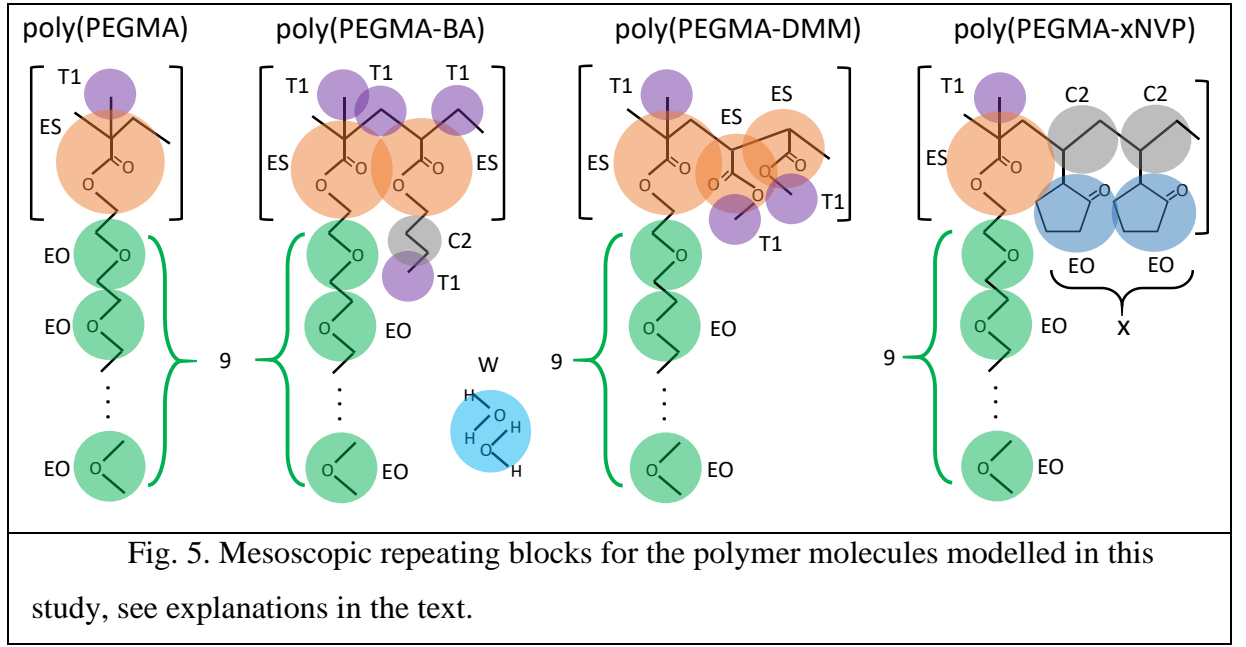

Fig. 5. Mesoscopic repeating blocks for the polymer molecules modelled in this study, see explanations in the text.

Here W stays for two water molecules, whereas C2, T2, EO and ES – for a $CH_2$, $CH_3$, glycol and ester group, respectively, as introduced in Ref [29]. Following the water solubility and estimated effective volume of a vinylpyrolidon (VP) group, it can be approximated by the same EO block as the glycol group. Similar approximate representations are assumed for the butyl acrylate (BA) and methyl metacrylate (MM) groups, see Fig.MF1. DPD simulations are performed in reduced units with the lengthscale given by the diameter of the W sphere, and the energy in units of $k_B T = 1$.

Each building block is interpreted as a soft sphere with its van der Waals diameter and an associate set of surface charges[29]. The latter are accounted for in an approximate way, by setting the respective strength $a_{ij}$ of the pairwise repulsion between $i$-th and $j$-th sphere:

$$F_{ij}^C = \begin{cases} a_{ij}\left(1 - \frac{r_{ij}}{R_{ij}}\right), & r_{ij} < R_{ij} \\ 0, & r_{ij} \geq R_{ij} \end{cases} \quad (4)$$

Here $F_{ij}^C$ stays for the amplitude of a conservative repulsive force, $r_{ij}$ is the distance between the centers of the spheres and $R_{ij}$ is the pairwise van der Waals. We use the same values for $R_{ij}$ and $a_{ij}$ as obtained in Ref.[29], these are reproduced from the latter source in Tabs.MT1 and MT2.

|    | W            | ES           | EO           | C2           | T1           |
|----|--------------|--------------|--------------|--------------|--------------|
| W  | 1.000[25.00] |              |              |              |              |
| ES | 1.071[22.53] | 1.141[22.00] |              |              |              |
| EO | 1.058[21.81] | 1.129[24.01] | 1.116[22.50] |              |              |
| C2 | 1.037[45.45] | 1.108[21.50] | 1.095[23.78] | 1.074[22.00] |              |
| T1 | 0.978[46.35] | 1.048[21.61] | 1.036[24.18] | 1.015[22.92] | 0.955[24.00] |

Tab.3. Effective pair interaction parameters $R_{ij}[a_{ij}]$ for the pairwise repulsive interaction (1) between spheres of different type, reduced DPD units. Reproduced from Ref.[Lavag2021].

Besides nonbonded forces (4), spherical particles are subject to the bonded forces within each molecule, these include bonding and bond angle forces, with the respective amplitudes $F_{ij}^B$ and $F_{ij}^A$ given by the expressions:

$$F_{ij}^B = -k_b(r_{ij} - b_{ij}), \qquad F_{ij}^B = -k_a(\theta_{ij} - t_{ij}) \qquad (5)$$

where, following[31,29], we used the values $k_b = 150$ and $k_a = 5$ in DPD units introduced above. The bond lengths $b_{ij}$ are evaluated according to the approximate recipe suggested in Refs.[32,29], where the $CH_2 - CH_2$ bond is estimated as $0.39$ and each extra heavy atom within a group contributes additional length of $0.1$, both in DPD units. This results in the following set of bond lengths used in this study: 0.39 (C2-C2), 0.55 (C2-ES), 0.49 (C2-EO), 0.29 (C2-T1), 0.79 (ES-ES), 0.45 (ES-T1), 0.69 (ES-EO), 0.59 (EO-EO).

Additionally, there are the pairwise friction and random forces, introduced for each pair of interacting spheres in a usual for the DPD simulations way[27]. Their respective amplitudes are given via:

$$F_{ij}^D = -\gamma w^D(r_{ij})(r_{ij} \cdot v_{ij})\frac{1}{r_{ij}}, \qquad F_{ij}^R = \sigma w^R(r_{ij})\theta_{ij}\Delta t^{1/2} \qquad (6)$$

where $r_{ij}$ is the radius-vector that connects particles $i$ and $j$, $v_{ij} = v_i - v_j$ is the difference between their respective velocities, $\gamma$ and $\sigma$ are the amplitudes of the dissipative (friction) and random forces, respectively, $w^D(r_{ij})$ and $w^R(r_{ij})$ define the decay of both forces with the separation between particles

$$w^D(r_{ij}) = [w^R(r_{ij})]^2 = \begin{cases} \left(1 - \dfrac{r_{ij}}{R_{ij}}\right)^2, & r_{ij} < R_{ij} \\ 0, & r_{ij} \geq R_{ij} \end{cases}$$

$\theta_{ij}$ is the Gaussian random variable. For more details see Ref. [27]. Because of a relatively high value of a spring constant $k_b = 150$, as compared to the initial value of $k_b = 4$ used in Ref.[27], the simulations are performed with a small time step of $\Delta t = 0.01^{29}$. Each simulated case comprises in total $4 \cdot 10^6$ DPD steps, of which the first $5 \cdot 10^5 - 10^6$ (depending on the polymer) are allowed for equilibration.

To match experimental conditions, one would prefer the situation, where a simulation box contains at least few micelles to allow averaging over their sizes and shapes. On a top of that, relatively long simulation times are required, because of slow dynamics of macromolecules and their sub-micellar aggregates, and, therefore, the metastability effects. Both factors put a huge strain on the expenditure of computational time, especially for the models with explicit solvent, as is the case for the DPD approach, and may turn the simulations completely unfeasible.

To overcome these difficulties, we restrict molecular masses of the model amphiphiles. Namely, all model molecules contain fixed number of 16 PEG side chain blocks in the case of each polymer, assembled according to Fig.5. Simulation box is also restricted in size to the $40 \times 40 \times 40$ in DPD units. Then, assuming the number density of 3, as first suggested by Groot and Warren[27] based on fits of the model compressibility to that of water, one arrives at the total number of spherical particles equal to 192000. Each molecule type is assembled according to Fig.5, therefore, we consider in total 6 model amphiphiles: poly(PEGMA), poly(PEGMA-co-BA), poly(PEGMA-co-DMM) and poly(PEGMA-co-xNVP), where x=2,3,4. This allows us to both study the role of the type of separators placed in between PEGMA side chains, and their number, in the case of NVP.

Following suggested protocol for simulation of micellization[33], there are several ways to define the micellar aggregates based on positions of individual monomers of the amphiphiles. We used the following criteria: two amphiphiles, $i$, $j$, are assumed to belong to the same aggregate if there are at least two pairs of monomers (where one monomer of a pair belongs to $i$th amphiphile and the other – to $j$th one) with the separation less than $1.5$ unit lengths. Then, the network of linked molecules is constructed, with the periodic boundary conditions being taken into account. This network defines all amphiphilic molecules participated in an aggregate and, consequently, the set of all monomers they are formed of. To collect statistics on aggregation, the coordinates of all monomers are saved after each $10^3$ DPD steps in a course of simulations and then postprocessed for the analysis.

We found that a simulation box of a chosen dimensions contains either sub-micellar aggregates only, or a single micelle with a few sub-micellar aggregates. If a micelle is present, its size and shape fluctuate in a course of simulations, hence we obtain the time averages for the micelle size and shape. We expect that these averages are close to the ones that might be obtained over an ensemble of micelles if more than one micelle would be present in a simulation box at each time instance. The latter case, however, is too costly to simulate directly for the polymers of the size considered here.

One of the problems that hamper simulation of micellization is a metastability[34]. It however, is much reduced in the DPD simulation approach, because of the possibility for polymer chains to cross one another easily, as a consequence of a soft nature of interaction potentials. In certain case this is serious shortcoming (e.g., lack of the entanglement effects affecting polymer dynamics), in another – a huge benefit (e.g., speeding up the microphase segregation driven phenomena). The latter effect is known since the very beginning of the DPD method[35,36]. Despite of this benefit of the DPD approach, micellization is still hampered by slow native dynamics of isolated macromolecular amphiphiles and their sub-micellar aggregates. Therefore, to make sure that we cover enough statistics in micelles sizes and shapes, we performed two micellization runs. Runs marked via (1) are started from a random arrangement of polymers within the simulation box, whereas run (2) comprises two stages. During the first stage, all monomers of the polymers are made temporarily hydrophobic, with the repulsion parameter $a_{ij} = 40$ in Eq. (4) and all chains are made fully flexible, $k_a = 0$ in Eq. (5). This promotes quick formation of a closely packed spherical blob, which is further equilibrated for $1 \cdot 10^5$ DPD steps. On the second stage, all the repulsion parameters are chosen according to the types of beads (see, Tab.3) and the chains are made semiflexible with $k_a = 5$. As the result, the blob expands and then either stays a single micelle (sometimes loosing some molecules), or splits into sub-micellar aggregates.

To compare the output of the runs (1) and (2), we monitor a characteristic property of the size $N_m^{max}$ for the largest aggregate (associated with a micelle) at each time instance. The results for the time evolution of $N_m^{max}$ are shown in Fig.6 for the case of $N = 80$ molecules for all 6 amphiphiles. One can see that after the system equilibrates during $5 \cdot 10^5 - 10^6$ DPD steps (depending on the polymer type), both approaches provide very similar time evolution for $N_m^{max}$. The time interval from $5 \cdot 10^5 - 10^6$ to $4 \cdot 10^6$ DPD steps provides the statistics for all the properties of the interest.

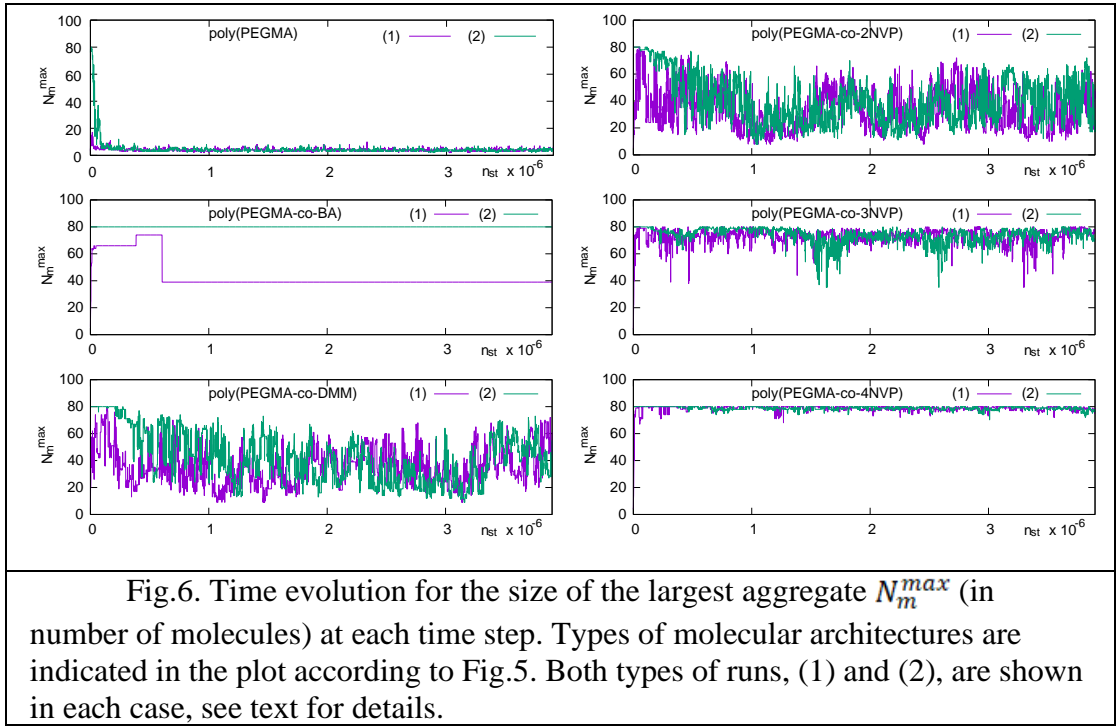

Fig.6. Time evolution for the size of the largest aggregate $N_m^{max}$ (in number of molecules) at each time step. Types of molecular architectures are indicated in the plot according to Fig.5. Both types of runs, (1) and (2), are shown in each case, see text for details.

At first, one needs to distinguish between the sub-micellar and micellar regimes. To this end, we evaluate the histograms representing the distributions $p(N_m)$ for the sizes $N_m$ of all aggregates found at each time instance. These distributions may, or may not (as pointed out in Refs.[33,29]) possess a reasonably deep minima between the sub-micellar and micellar peaks. The results for $p(N_m)$ are collected in Fig.7 for the cases of all 6 polymers and all solute densities being considered, the latter are expressed via the number of solute molecules $N$. We interpret these plots in a following way. The polymers poly(PEGMA), poly(PEG-co-DMM) and poly(PEG-co-2NVP) demonstrate a single peak at small $N_m$ only, which is associated with a sub-micellar regime. This is valid in the whole interval of solute densities considered, $N = 10 - 80$, which translates into the volume fractions of a solute $f = 0.016 - 0.128$. For the polymers with a higher number of NVP groups, the second peak develops at higher densities, $N = 60, 80$, which is located at the aggregate sizes $N_m$ approaching $N$. These peaks indicates the micellar regime. The critical micelle concentration (CMC) may, therefore, be roughly estimated as $N_{cmc} > 80$ ($f_{cmc} > 0.1$) for poly(PEGMA), poly(PEGMA-co-DMM) and poly(PEGMA-co-2NVP, $N_{cmc} \approx 60$ ($f_{cmc} \approx 0.086$) for PEG-3NVP, and $N_{cmc} \approx 40$ ($f_{cmc} \approx 0.064$) for PEG-

4NVP. Such two-maxima form for the $N_m$ distributions bears similarities with the first-order phase transitions, e.g. a liquid-vapor of nematic-isotropic ones[37]. Unfortunately, the simulations are not able to analyze the micellization of the poly(PEGMA-co-BA), as the model system is found to be trapped in a long living metastable state, which strongly depends on the initial arrangement of the polymers within a simulation box. For runs of type (2) for this polymer, displayed in Fig.7, the histograms for $p(N_m)$ turned into a δ-like spikes $p(N_m) = \delta(N_m - N)$.

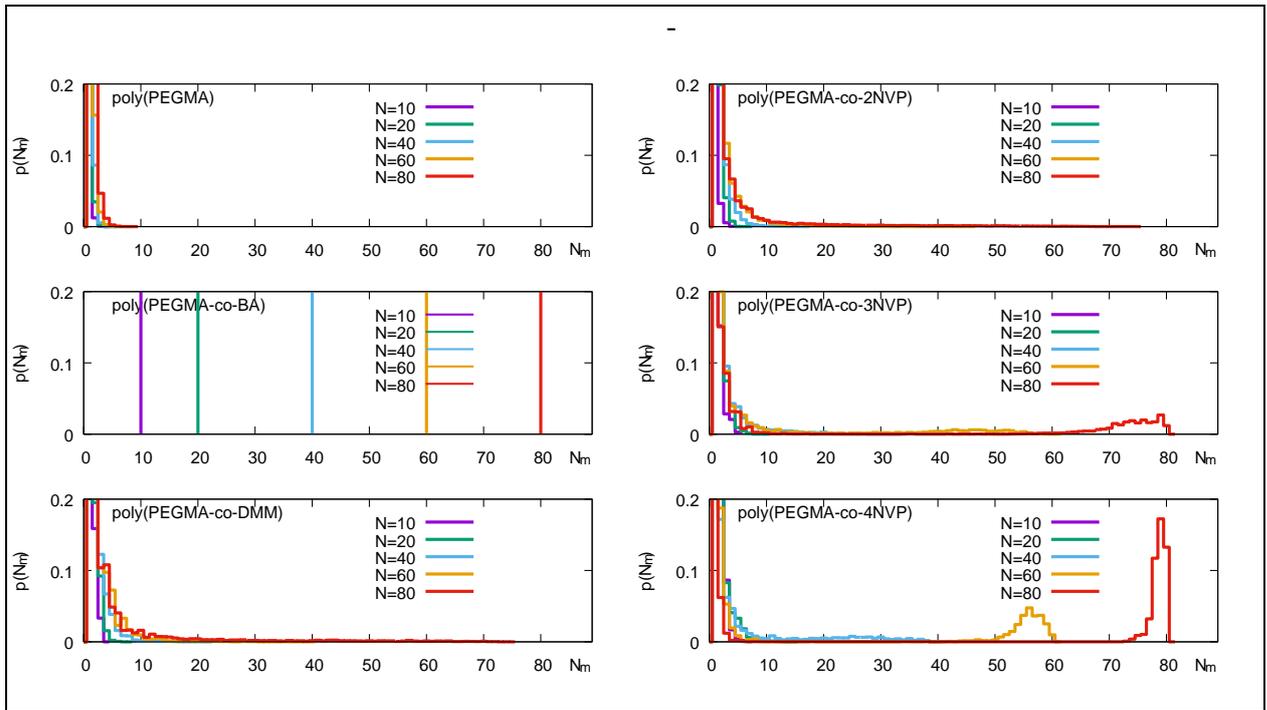

Fig. 7. Histograms representing $p(N_m)$ distributions for the size of all aggregates in a solution for each molecular architecture at each solute density represented in terms of the number of solute molecules $N$.

It is expected that the analysis of the size $N_m^{max}$ for the largest aggregate only (the "giant component" in the terminology of the networks theory) will separate the sub-micellar and micellar regimes more clearly. This is found to be the case, as indicated by the histograms for the distribution $f(N_m^{max})$ shown in Fig.8. Each histogram is characterized by a single maximum with its location either at small $N_m^{max}$ (sub-micellar regime) or at $N_m^{max}$ which approaches $N$ (micellar regime). The CMC is characterized here by a low amplitude and broadly spread shape for the $f(N_m^{max})$ with its maximum located at $N_m^{max} \approx N/2$. Basically, the same CMC estimates hold, as found previously in Fig.7. More precise estimates for CMC require performing simulations with a finer grid in $N$.

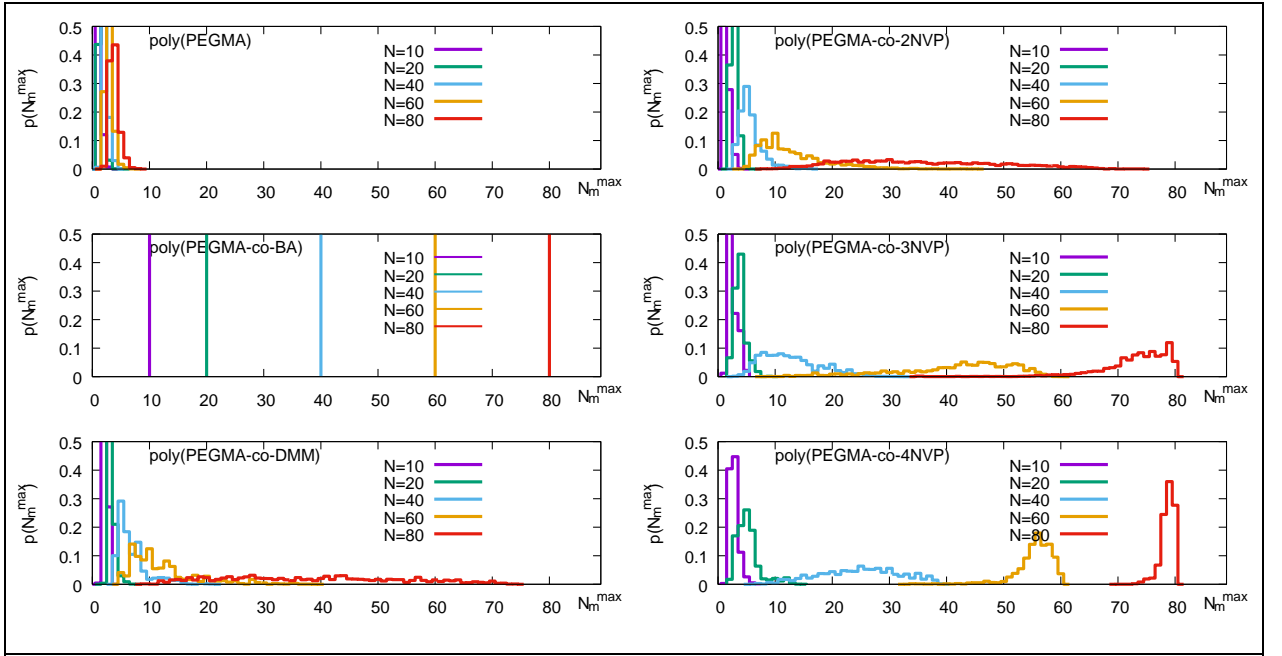

Fig. 8. Histograms representing $p(N_m^{max})$ distributions for the size of the largest aggregate in a solution for each molecular architecture at each solute density represented in terms of the number of solute molecules $N$.

The mean aggregation number is evaluated according to a standard equation

$$\langle A_N^{max} \rangle = \frac{\langle N^2 \rangle}{\langle N \rangle}, \qquad (7)$$

where the averaging is performed over the $p(N_m)$ distribution in each case. The plot reflecting the behavior of the mean aggregation number on the polymer concentration is shown in Fig.9 for each type of a polymer. For the sake of convenience, we plot the normalized property, $\frac{\langle A_n^{max} \rangle}{N}$ vs volume fraction of the polymer $f$. The value $\frac{\langle A_n^{max} \rangle}{N} = 1$ means that a micelle contains all available molecules in a solution and no sub-micellar aggregates are present. This is the always case for the poly(PEGMA-co-BA) but, in our opinion, this result might be artificial, due to extremely long-lived metastable states and indicates that the model for this polymer requires an essential refinement. This is a bit unlucky, as this particular polymer demonstrates the highest hydrolysis efficiency from the entire group (see Fig.2). For the other polymers, one has for $\frac{\langle A_n^{max} \rangle}{N}$ a curve that can be fitted well by the logistic function

$$\frac{\langle A_n^{max} \rangle}{N}(f) = 1 - C + \frac{C}{1+\exp\left[-(f-f_0)/k\right]} \qquad (8)$$

where $C$ is a characteristic constant for each type of a polymer. At $f \to 1$ and $f_0 \ll 1$ there is always a single micelle, $\frac{\langle A_n^{max} \rangle}{N} \to 1$, as this case represents a solvent-free tight packing of the polymer particles within a simulation box of a finite size. The value $f_0$ is the position for an inflection point for the $\frac{\langle A_n^{max} \rangle}{N}(f)$ curve and is found to be surprisingly close to the previous estimates for the CMC $f_{cmc}$, based on Figs. 7 and 8. Namely, $f_0 \approx 0.103$ (vs $f_{cmc} > 0.1$) for poly(PEGMA-DMM), $f_0 \approx 0.12$ (vs $f_{cmc} > 0.1$) for poly(PEGMA-2NVP), $f_0 \approx 0.078$ (vs $f_{cmc} \approx 0.086$) for poly(PEGMA-3NVP), and $f_0 \approx 0.062$ (vs $f_{cmc} \approx 0.064$) for poly(PEGMA-4NVP). This is not surprising, as, if one would trace the maxima positions for the sub-micellar (gas-like) and micellar (liquid-like) regimes in Fig.7 separately, this will result in two separate branches – one for each regime, with the discontinuous jump between both, reminiscent of the first order phase transition[37]. The position of the jump is associated with the phase transition point. However, when averaging over the whole distribution $p(N_m)$, one obtains a single smooth curve, where the phase transition point is close to the inflection point of the curve. Therefore, the latter point, $f_0$ can be associated as an estimate for the CMC. One may conclude the results shown by Fig.9 by observing, that separation of the PEG chains by any intermediate chains-inserts, such as DMM, BA, or NVP lead to reduction of the CMC. This is especially clear from comparison of the cases for poly(PEGMA-co-xNVP) (see Fig.9, r.h.s.), where the increase of $x$ (making stronger separation between the PEG chains) leads to both essential increase of the average aggregate number at given concentration $f$, and to reduction of the CMC value. This confirms experimental findings made in section 2.

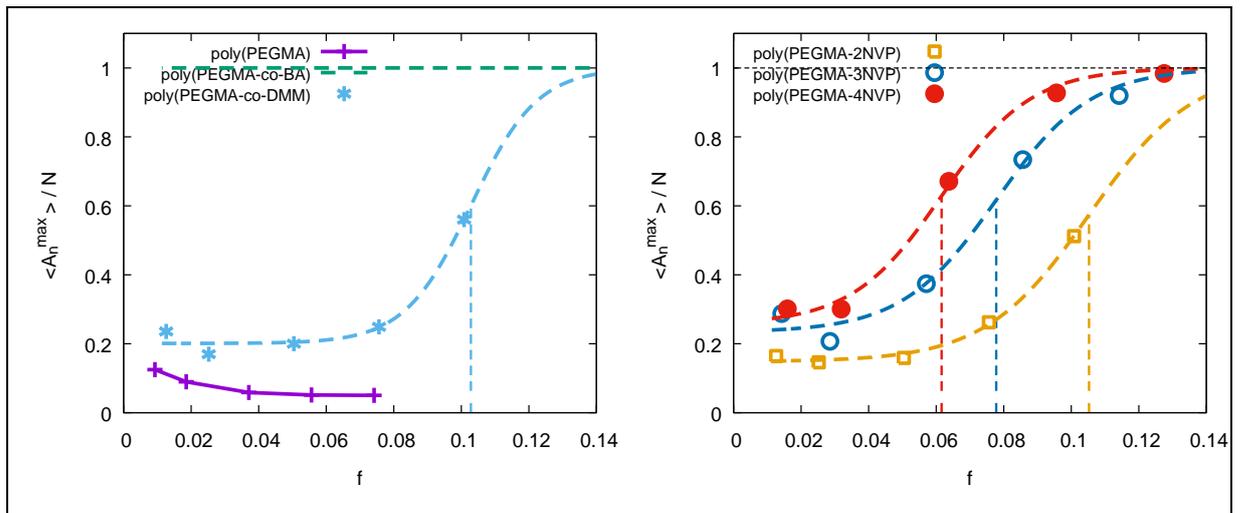

Fig. 9. Normalized mean aggregation number $\langle A_n^{max} \rangle / N$ vs volume fraction $f$ for each polymer type (indicated within a plot) as a function of the polymer volume fraction $f$. Verical dashed lines indicate inflction points with their values found to be very close to the estimated CMC values.

The hydrodynamic radius for each aggregate is evaluated according to the expression

$$R_h^{-1} = \langle 1/r_{ij}\rangle_{i\neq j} \quad (9)$$

where $i, j$ run over all the particles of all polymer molecules that belong to a given aggregate (with the periodic boundary conditions accounted for). Fig.10 shows histograms for the distributions $f(R_h^{max})$ of the values of hydrodynamics radius $R_h^{max}$ for the largest aggregate. One can see that the shapes of $f(R_h^{max})$ looks very similar to their counterparts for $f(N_m^{max})$ shown above in Fig.8. In particular, $f(R_h^{max})$ is characterized by a clear maximum in both sub-micellar (at low $R_h^{max}$) and micellar (at high $R_h^{max}$) regimes and has a broad shape that is spread over large interval of values when close to CMC.

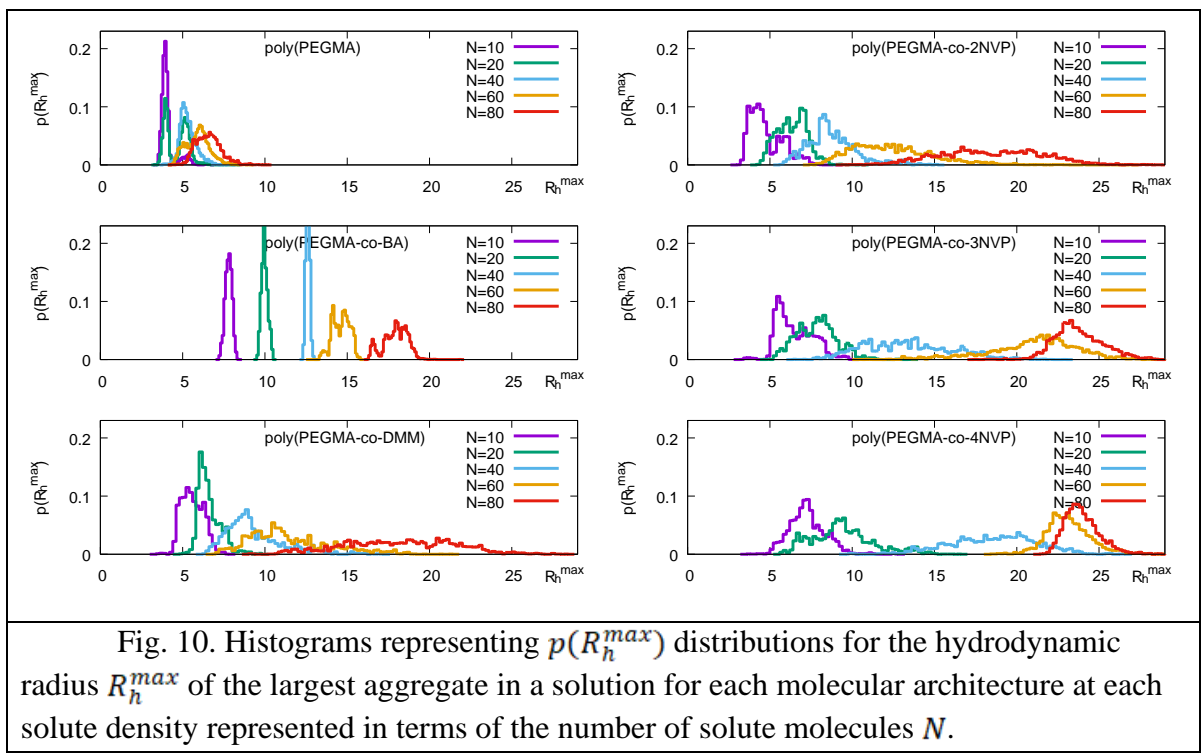

Fig. 10. Histograms representing $p(R_h^{max})$ distributions for the hydrodynamic radius $R_h^{max}$ of the largest aggregate in a solution for each molecular architecture at each solute density represented in terms of the number of solute molecules $N$.

The plots for the average $\langle R_h^{max}\rangle$ are shown in Fig. 11. At $f \to 1$ the value of $\langle R_h^{max}\rangle$ approach certain maximum value of about 25 reflecting the finite dimensions of a simulation box. As one can see, the behavior of this property for the case of poly(PEGMA-BA) polymer differs drastically from the others, demonstrating no infliction point at $f > 0.01$. Fitting procedure can be done when one assumes extremely small value for the $f_0 < 0.01$ in this case, shown in the figure. The conclusions arising from Figs.10 and 11 recall these, that were made earlier for the average aggregation number. In particular, the separation of the PEG chains by any of DMM, BA, or NVP groups, lead to the increase of the average hydrodynamics radius. We also note, that the effect is stronger for more hydrophobic separating groups, such as BA.

Unfortunately, the direct comparison for the hydrodynamics radius against the experimental values is not straightforward due to several reasons. Firstly, there is considerable difference in sizes between experimental and model molecules ranging from the factor of $17$ for poly(PEGMA) to the factor of $2$ for the poly(PEGMA-co-NVP) polymer in terms of model particles shown in Fig.5. Given the specific architectures of all molecules considered here, they will obey rather complex scaling laws that will require a separate study, e.g. by modeling a homologous series of molecules with different sizes and extrapolating the results to the sizes comparable to the experimental systems. Secondly, the copolymerization is a random process and produce a set of different molecular architectures that are characterized by statistically similar composition only, and not by exactly the same molecular architecture, which coincides with its model counterparts shown in Fig.5. Finally, the model contains a number of simplifications in terms of hydrogen bonding abilities and average electrostatic properties of its constituents, and these may be crucial for faithful reproduction of micellization.

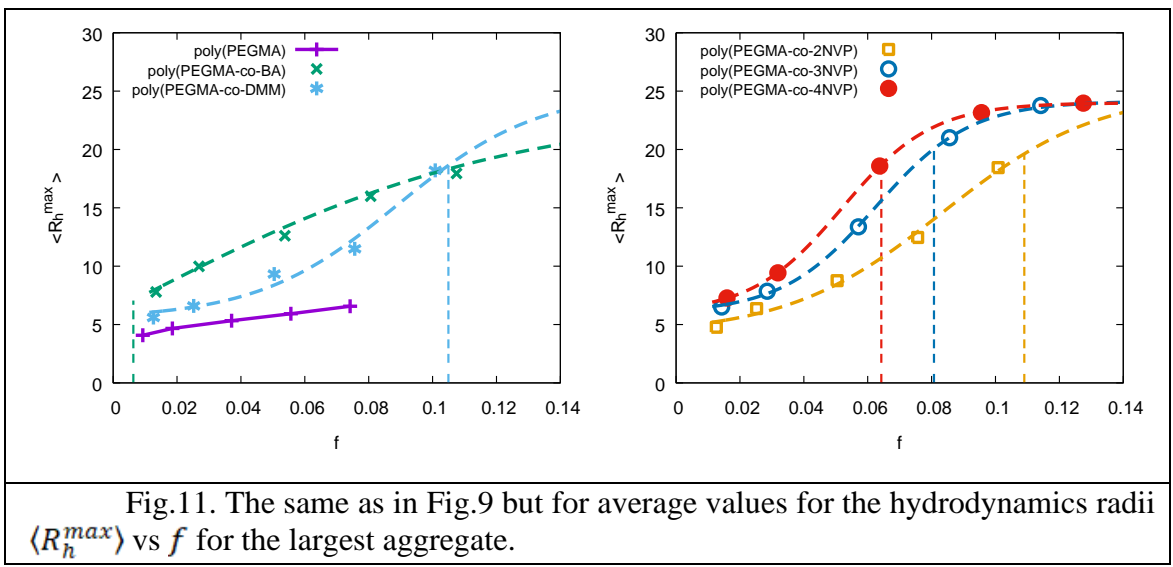

Fig.11. The same as in Fig.9 but for average values for the hydrodynamics radii $\langle R_h^{max} \rangle$ vs $f$ for the largest aggregate.

Another important property of interest is the shape of the micelle. To distinguish between prolate, spherical and oblate shapes we use the so-called shape descriptor $S_f$ defined as

$$S_f = -\frac{\lambda_2 - \langle \lambda \rangle}{\langle \lambda \rangle} \qquad (10)$$

where $\lambda_1 \geq \lambda_2 \geq \lambda_3$ are the eigenvalues of the gyration tensor evaluated for the micelle, where the latter is interpreted as a collection of centers of mass of each individual particle (for more details, see Ref.[38] and references therein), and $\langle \lambda \rangle = (\lambda_1 + \lambda_2 + \lambda_3)/3$ is their average value. This property is convenient due to its symmetry: it is equal to zero for a spherical shape, is $-1$ for infinitely flat disc and is $+1$ for the infinitely thin long rod. All intermediate oblate and prolate shapes fall in between this interval of values. The plots for the average $\langle S_f^{max} \rangle$ are shown in Fig. 11. One can see, that for all cases of polymer molecules, except the poly(PEGMA-co-BA), the shape factor is positive monotonically decays towards zero with the increase of the polymer concentration $f$. This behavior indicates a prolate micellar shape at low concentrations $f$ (presumably, due to semi-rigidity of the polymer backbones), which gradually became more spherical with the increase of the concentration (when the micellar size became compatible with the average dimensions of individual polymers and the latter are able to assemble into more spherical objects favored by the free energy considerations). This is in an agreement with the experimental findings shown in Fig.4.

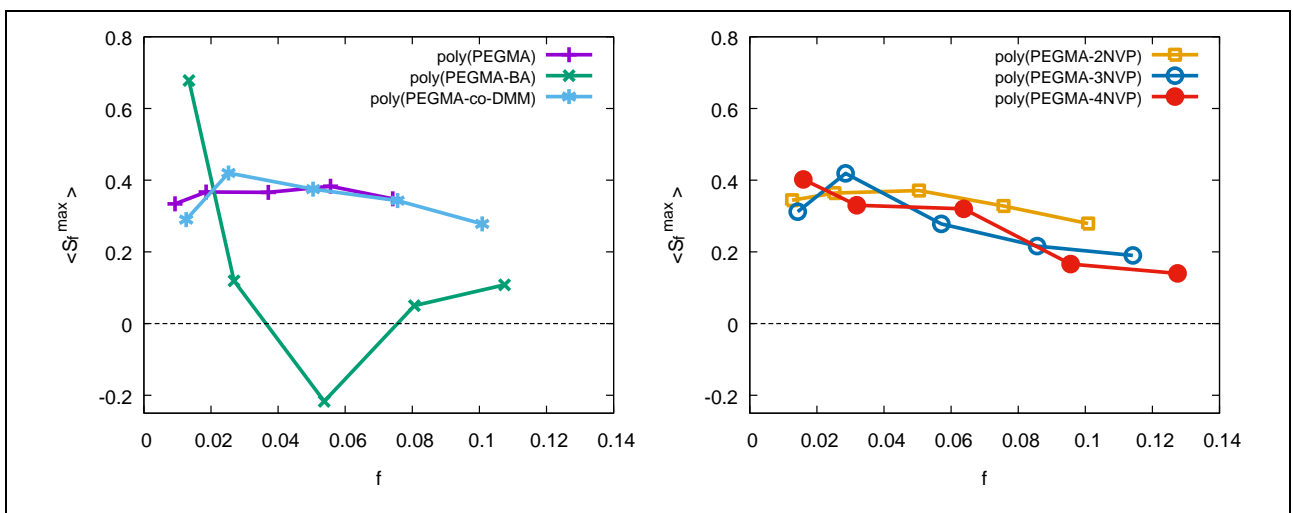

Fig.12. Averages values for the shape factor $\langle S_f^{max} \rangle$ vs $f$ for the largest aggregate shown for all polymers as functions of the polymer volume fraction $f$. The lines are only the guides for eyes that merely connect data points.

The case of the poly(PEGMA-co-BA) polymer is a special one, as the micelles demonstrate extremely extended, rod-like shape with $\langle S_f^{max} \rangle \sim 0.7$ at lowest concentration $f \sim 0.01$, and with the increase of $f$ may be found in both prolate and oblate (negative values for $\langle S_f^{max} \rangle$) shape (see Fig.12). It is not quite clear yet for us whether this is the result of the metastability issue, mentioned for this polymer above, or some inherent feature of this particular polymer.

Visualization of model micelles is presented in Fig.13. In most cases, micelles have an appearance of small networks, held together by a small number of links between hydrophobic groups of adjacent backbones. The case of poly(PEGMA-co-BA) is different and is characterized by double strands, formed by adjacent backbones merged side-by-side. These micelles have more compact structures. This, however, might be the artefact of the initial configuration preparation and require a further study. Nevertheless, all structures are relatively porous and, geometrically, are able to hold a large number of guest enzymes. This will be the subject of the subsequent study.

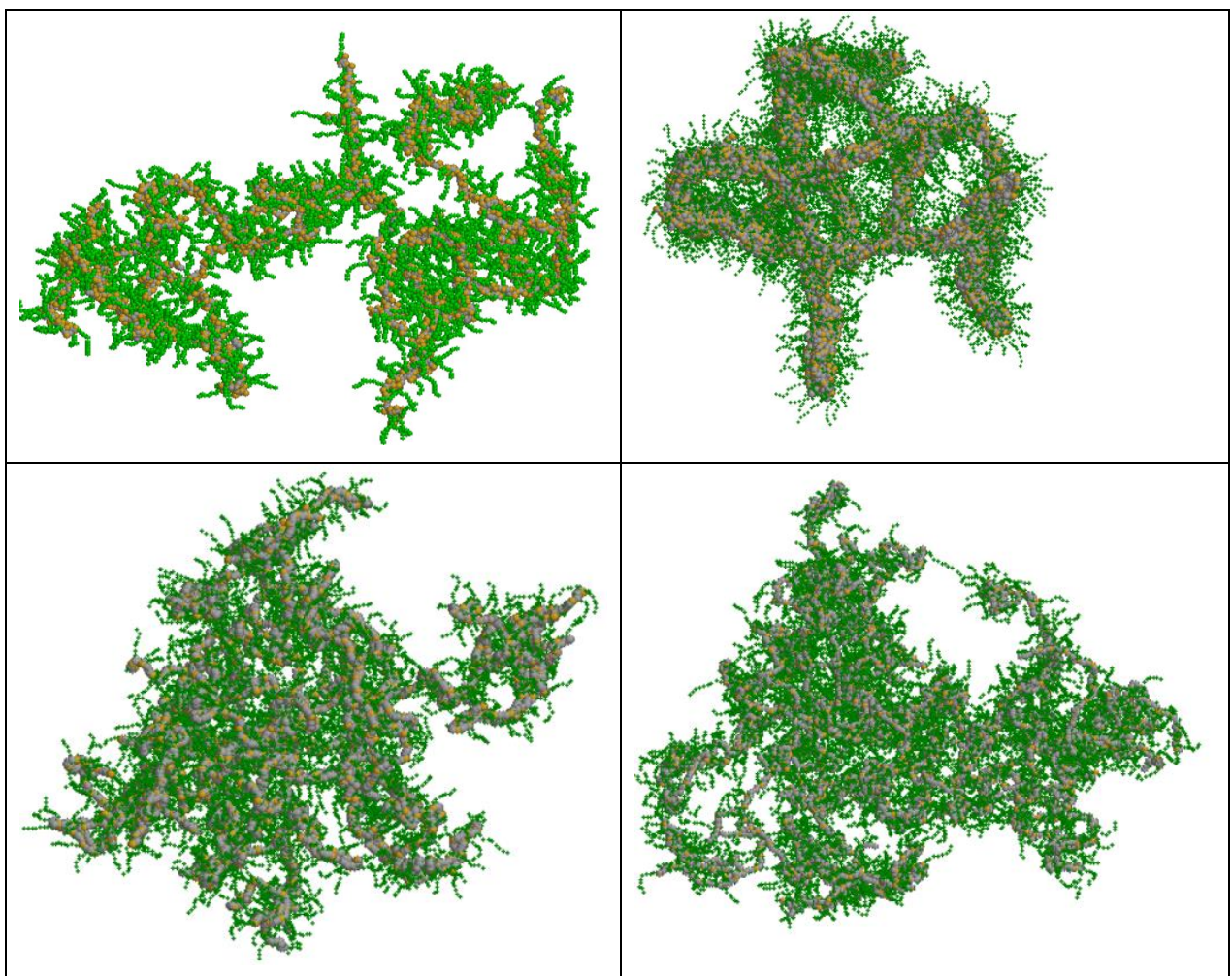

Fig.13. Snapshots of micelles, assembled from 80 polymers in water solutions. Top row: poly(PEGMA-co-DMM) (l.h.s.) and poly(PEGMA-co-BA) (r.h.s.). Bottom row: poly(PEGMA-co-3NVP) (l.h.s.) and poly(PEGMA-co-3NVP) (r.h.s.). Hydrophobic PO particles (green) are showed in smaller size to allow more clear view of the ES particles (yellow) and hydrophobic C2 and T1 particles (gray).

**Conclusions**

This paper combines experimental and computer simulation studies to examine the properties of the aggregates (micelles) formed by a set of the synthesized PEG - copolymers. The main focus of the work is to answer the question of whether or not such a class of synthetic polymers can be a good candidate as a host system for enzymes to mimic natural cellulosomes for efficient cellulose hydrolysis. To this end, the tests of the cellulose depolymerization reported here in Fig.2 demonstrated some improvement for the process as compared to the case of the mixture of free enzymes. The study was directed towards the search of possible links between the micellization of each polymer structure and the hydrolysis efficiency of the host-guest system based on its micellar structure. In particular, it was found that poly(PEGMA-co-BA) micellar structures are characterized by the highest glucose production rate. The molecular architecture, specifically the separation between PEG side chains along the polymer backbone, is beneficial for micelle formation. The micellar structures of this copolymer are stabilized by strong backbone-backbone interactions and possess pocket-like domains that can accommodate enzyme molecules. All these results call for future refinement of the molecular models and steer research towards looking for the polymers with this given set of properties of their micellar structures.

## Acknowledgements

The authors acknowledge financial support from the CRDF Global award 66705.